\documentclass[journal]{IEEEtran}
\usepackage{cite}

\ifCLASSINFOpdf
\else
\fi
\usepackage{url}

\usepackage{tabularx}
\usepackage{float}
\usepackage{graphicx}
\usepackage{nomencl}
\usepackage{amsmath}
\usepackage{booktabs}
\usepackage{stfloats}
\usepackage{amssymb}
\usepackage{multirow}
\usepackage{subfigure}
\usepackage{xcolor}

\begin{document}
%
\title{
A Smart Switch Configuration and Reliability Assessment Method for Offshore Wind Farm Electrical Collector System

}
%
%
%

\author{Xiaochi~Ding,~\IEEEmembership{Student Member,~IEEE,} Xinwei~Shen,~\IEEEmembership{Senior~Member,~IEEE,}
Qiuwei~Wu,~\IEEEmembership{Senior~Member,~IEEE,} Liming~Wang,~\IEEEmembership{Senior~Member,~IEEE,} Dechang~Yang,~\IEEEmembership{Member,~IEEE} 
\thanks{Xiaochi Ding and Qiuwei Wu are with Tsinghua-Berkeley Shenzhen Institute, Tsinghua Shenzhen International Graduate School (SIGS), Tsinghua University. Xinwei Shen and Liming Wang are with Institute for Ocean Engineering, Tsinghua SIGS, Tsinghua University. Dechang Yang is with College of Information and Electrical Engineering, China Agricultural University. This work is supported in part by National Natural Science Foundation of China (No. 52007123) (Corresponding author: Xinwei Shen, sxw.tbsi@sz.tsinghua.edu.cn).
}
}

%
%

\markboth{}%
{Shell \MakeLowercase{\textit{et al.}}: Bare Demo of IEEEtran.cls for IEEE Journals}
%



\maketitle

\begin{abstract}

With the development of offshore wind farms (OWFs) in far-offshore and deep-sea areas, each OWF could contain more and more wind turbines and cables, making it imperative to study high-reliability electrical collector system (ECS) for OWF. Enlightened by active distribution network, for OWF, we propose an ECS switch configuration that enables post-fault network recovery, along with a reliability assessment (RA) method based on optimization models.
It can also determine the optimal normal state and network reconfiguration strategies to maximize ECS reliability. Case studies on several OWFs demonstrate that the proposed RA method is more computationally efficient and accurate than the traditional sequential Monte-Carlo simulation method. Moreover, the proposed switch configuration, in conjunction with the network reconfiguration strategy and proper topology, provides significant benefits to ECS reliability.


\end{abstract}


\vspace{-1em}
\section*{Nomenclature} 
\subsection*{Indices and Sets} 
\addcontentsline{toc}{section}{Nomenclature}
\begin{IEEEdescription}[\IEEEusemathlabelsep\IEEEsetlabelwidth{$i, j, x, y$}] 
\item[$br^f$] Index for the cable connected to feeder $f$.
\item[$f$] Index for feeders.
\item[$i, j, r, s, k$] Indices for nodes.
\item[$ij, rs$] Indices for cables.
\item[$u$] Index for fault events.
\item[$\mathrm{TS}/\mathrm{RS}$] Index for tripped stage / recovery stage.
\item[$\Psi_N/\Psi_N^{WT}$] Set of nodes / wind turbine nodes.
\item[$\Psi_C/\Psi_F$] Set of cables / feeders.
\item[$\Psi_i$] Set of nodes connected to node $i$.
\item[$\Psi_I^B/\Psi_I^S$] Set of cables with breaker/switch at the left end.
\item[$\Psi_J^B/\Psi_J^S$] Set of cables with breaker/switch at the right end.
\vspace{-2.5em}
\end{IEEEdescription}
\subsection*{Parameters} 
\addcontentsline{toc}{section}{Nomenclature}
\begin{IEEEdescription}[\IEEEusemathlabelsep\IEEEsetlabelwidth{$i, j, x, y$}] 
\item[$b_{ij}^{i/j,NO}$] Connection status of circuit breaker at $i/j$ on cable $ij$ under normal operation.
\item[$c_{CB}/c_{SW}$] Price of circuit breaker / isolation switch.
\item[$h_{ij}^f/h_k^f$] Cable-feeder/Node-feeder affiliation, 1 denoting cable $ij$ / wind turbine $k$ supplies power to the offshore substation through feeder $f$.
\item[$n_{CB}/n_{SW}$] Number of circuit breakers / isolation switches.
\item[$r/t$] Discount ratio / Operation time of the project.
\item[$u_d$] Wind turbine annual effective utilization hours.
\item[$B_{ij}$] Susceptance of cable $ij$.
\item[$EENT_0$] Expected energy not transmitted of the system without any breakers or switches.
\item[$M$] Big-M constant.
\item[$P_f^C/P_{ij}^C$] Power transmission capacity of feeder $f$/cable $ij$.
\item[$P_k/R_k$] Sent power / Rated capacity of wind turbine $k$.
\item[$\alpha$] Unit-price of offshore wind energy (\$/kWh).
\item[$\lambda_{ij}/\lambda_{k}$] Failure rate of cable $ij$ / wind turbine $k$.
\item[$\tau^{SW}/\tau^{RP}$] Time required to isolate / repair the cable fault.
\item[$\tau^{WT}$] Time required to repair the wind turbine fault.
\end{IEEEdescription}
\vspace{-2.5em}
\subsection*{Variables} 
\addcontentsline{toc}{section}{Nomenclature}
\begin{IEEEdescription}[\IEEEusemathlabelsep\IEEEsetlabelwidth{$i, j, x, y$}] 
\item[$b_{ij}^{i,rs}/b_{ij}^{j,rs}$] Connection status of breaker at $i/j$ at tripped stage after cable $rs$ fails.
\item[$f_{i j}^{rs, \mathrm{TS/RS}}$] Virtual fault flow variables, equal to 1 when virtual fault flows through cable $ij$ at tripped / reconfiguration stage after cable $rs$ fails.
\item[$f_{i}^{rs, \mathrm{TS/RS}}$] Virtual fault flow variables, equal to 1 when virtual fault flows through node $i$ at tripped / reconfiguration stage after cable $rs$ fails.
\item[$m_k^{rs}$] Fault impact variable, 1 when wind turbine $k$ is affected by the failure of cable $rs$.
\item[$n_k^{rs}$] Fault continuation variable, 1 when wind turbine $k$ still cannot send power after reconfiguration.
\item[$s_{ij}^{NO}/s_{ij}^{rs}$] Connection status of cable $ij$ in normal operation / after reconfiguration following fault of cable $rs$.
\item[$s_{ij}^{i,rs}/s_{ij}^{j,rs}$] Connection status of isolation switch at $i/j$ at reconfiguration stage after cable $rs$ fails.
\item[$C_{rel}$] Reliability-related cost of the offshore wind farm.
\item[$EENT$] Expected energy not transmitted of the system.
\item[$P_f^{rs}/P_{ij}^{rs}$] Power flowing through feeder $f$ / cable $ij$ after reconfiguration due to fault of cable $rs$.
\item[$P_k^{rs}$] Wind power sent by wind turbine $k$ after reconfiguration following fault of cable $rs$.
\item[$TID_k$] Turbine interruption duration of node $k$.
\item[$TIF_k$] Turbine interruption frequency of node $k$.
\item[$V$] Switch configuration comprehensive benefits. 
\item[$\theta_i^{rs}$] Voltage phase of $i$ when cable $rs$ fails.

\end{IEEEdescription} 
%
\IEEEpeerreviewmaketitle

\vspace{-1em}
\section{Introduction}
%
%
%
%
\IEEEPARstart{W}{ind} power is currently one of the fastest-growing forms of renewable energy. 
Offshore wind power has several advantages over onshore wind power, including higher wind speeds, higher annual utilization hours, and the ability to conserve land resources. As a result, there is potential for significant expansion in offshore wind power. The European Commission has projected that offshore wind power capacity will reach 450 GW by 2050 \cite{freeman2019our}.
However, as offshore wind farms (OWFs) continue to grow in size, more cables are required to connect the wind turbines (WTs) to the grid, increasing the vulnerability of their electrical collector systems (ECSs) \cite{berzan2011algorithms}. Submarine cables are located under the seabed, making their maintenance and repair extremely hard. Consequently, the mean time to repair (MTTR) may exceed two months \cite{zuo2020optimal}. The faults of WTs and cables will impact the electricity output of OWFs, leading to significant economic losses. Hence, it is necessary to research not only how to optimize ECS topology but also its reliability assessment (RA) methods \cite{hou2019}.\par
Reliability analysis of power systems has been widely studied in the literature. Ref \cite{allan2013reliability} discusses the definitions and calculation methods of reliability indices, such as expected energy not supplied (EENS), loss of load probability (LOLP), etc. Early research on RA mainly focused on distribution networks\cite{li2002incorporating,atwa2009reliability,heydt2010distribution,munoz2016reliability,munoz2017distribution,li2019analytical,li2020reliability}. However, with the increase in size of OWFs, RA has become indispensable for ensuring their economic and reliable operation.
Currently, RA methods can be divided into time-series simulation methods and analytical methods. References
\cite{wu2017economics,paul2019new,han2012four} use Monte-Carlo simulation to evaluate the reliability of OWFs or power systems containing wind farms. However, applying the simulation method to calculate the reliability of large-scale OWFs requires generating thousands of Monte-Carlo chronological sample states, which may greatly reduce computational efficiency.\par

Various analytical methods have been studied for evaluating the reliability of offshore wind farms (OWFs) \cite{dahmani2014reliability, dahmani2017, huang2010reliability, zhao2007, zhao2009, abeynayake2021, wang2018}. One of them is based on reliability block diagram and minimal path techniques to calculate reliability indices such as EENS and annual outage hours \cite{dahmani2014reliability, dahmani2017}. However, efficiency becomes an issue when the system becomes large, so two approximate calculation methods have been presented in \cite{dahmani2014reliability}. Another approach treats the WT string as an equivalent generator and applies traditional reliability evaluation methods for the electric system to study the availability of OWFs \cite{huang2010reliability}. References \cite{zhao2007, zhao2009} propose the index of generation ratio availability and its analytical calculation method to assess the performance of the OWF's ECS. Ref
\cite{abeynayake2021} adopts a method combining multi-state Markov process and universal generating function to assess system reliability.
Furthermore, case studies demonstrate the necessity of considering ECS's reliability in OWF availability assessment.
In addition, some scholars have attempted to consider reliability in wind farm planning problems \cite{wu2017economics, paul2019new, dahmani2017, zhao2009}. In\cite{paul2019new}, a bi-level optimization model is proposed and sequential Monte-Carlo simulation has been applied at the lower level to assess the reliability of the Pareto-front solutions (of wind farm layouts) obtained at the upper level. Refs \cite{dahmani2017} and \cite{zhao2009} use the genetic algorithm to optimize ECS topology and calculate the reliability of optimization results analytically.\par

\begin{figure}[!htbp]
\centering
\includegraphics[width=3.39in]{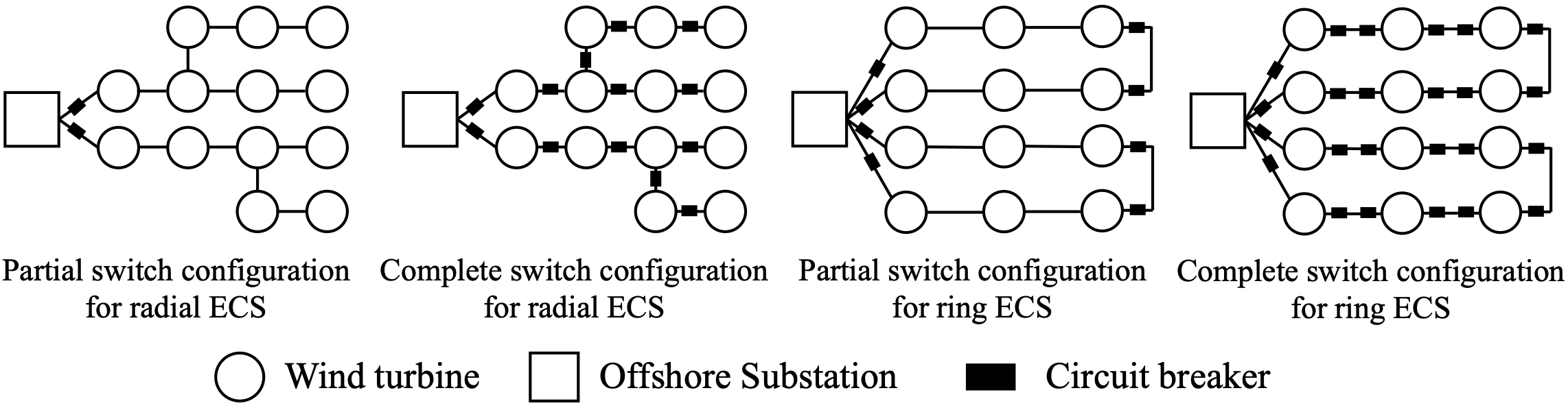}
\vspace{-0.8em}
\caption{Traditional ECS switch configurations \cite{Tan2013, Chen2019}. }
\label{Config}
\vspace{-0.5em}
\end{figure}

Various methods have been proposed for RA, but most of them focus on radial ECS with a single substation and often assume traditional switch configurations, as shown in Fig. \ref{Config}, or even ignore them. Although refs \cite{sannino2006,wang2018} have partly addressed these issues, detailed switch deployment is still not considered. As OWFs move towards larger-scale development, ECSs with multiple substations, complex topologies, and flexible switch configurations are becoming a trend, and the above research may not be suitable for future ECSs. Additionally, most methods cannot be integrated into the planning process, as they can only be used as a posterior simulation step to check if reliability requirements are met, which may result in unnecessary over-investment. Therefore, a reliability assessment model that can be embedded in the ECS planning process needs to be proposed.\par
\vspace{-0.7em}
It is worth noting that, the similarities between power distribution systems (PDSs) and ECSs are obvious\cite{shen2021large}. The PDS has been studied for decades, and RA models that can be integrated into the PDS's planning have been developed\cite{munoz2016reliability,munoz2017distribution,li2019analytical,li2020d,li2020reliability,wang2023}. In these models, reliability requirements could be considered as constraints in the planning stage.\par 
In this paper, we propose a smart switch configuration, along with a RA method for ECSs that draws upon evaluation theories and ideas from PDSs. The contributions have been summarized below:\par

\begin{itemize}
\item[1)]
Devise a smart switch configuration for OWF's ECS, which can support post-fault network reconfiguration at a relatively low cost, thereby enhancing the reliability and economy of the ECS.

\item[2)] 
Correspondingly, propose a RA model applicable to ECS with the smart switch configuration. It offers higher computation efficiency than Monte-Carlo method. Besides analyzing the reliability of existing ECSs, it can also be integrated into ECSs planning model. The potential applicability is demonstrated in the paper.
\item[3)] 
Develop another RA model for ECS considering detailed switch deployment based on virtual fault flow, and conduct a comprehensive benefits analysis and comparison for different switch placement strategies.

\end{itemize}\par

The remainder of this paper is arranged as follows. We introduce the conceptional analysis in Section II. The mathematical models are formulated in Sections III and IV. We present numerical tests in Section V, followed by discussions in Section VI. Section VII concludes the paper.\par

\section{Conceptional Analysis}
\subsection{Reliability Indices Calculation}
To describe the system reliability, the reliability indices should be discussed first. In this paper, we use \textit{expected energy not transmitted} (EENT, unit: MWh) as a metric to characterize the overall reliability of the ECS. EENT's calculation depends on the node reliability indices, namely, \textit{turbine interruption frequency} (TIF) and \textit{turbine interruption duration} (TID).\par
Conventional expressions used to calculate TIF and TID are:\par
\begin{equation}
    TIF_i=fr_i
\end{equation}
\begin{equation}
    T I D_{i}=\sum_{j=1}^{fr_{i}} t_{i}^{j}
\end{equation}
where $fr_i$ represents the number of times the $i$th WT stops supplying power, $t_i^j$ represents the duration of each power outage, $j=1,2,\dots,fr_i$. 
The above indices calculation approach requires known historical data. In the absence of historical data, RA should be conducted with a probability and statistical approach to obtain the expected value of reliability indices. To achieve this, we define the contingency set containing WT and cable outages, and analyze the probability and the impact of each outage. 
We introduce binary variables $m_i^u$ to indicate whether $i$th WT is affected by the fault event $u$ and cannot generate power, and similarly, binary variables $n_i^u$ to indicate whether $i$th WT is still unable to transmit power after the network reconfiguration in the fault event $u$. With the introduction of these variables, the conventional calculation expressions (1) and (2) can be rewritten as follows:\par
\begin{equation}
T I F_{i}=\sum_{u} \lambda_{u} m_{i}^{u}
\end{equation}
\begin{equation}
T I D_{i}=\sum_{u} \lambda_{u}\left(\tau^{SW}_{u} m_{i}^{u}+\tau^{RP}_{u} n_{i}^{u}\right)
\end{equation}
where $\lambda_{u}$ denotes the probability of fault event $u$, while $\tau^{SW}_{u}$ and $\tau^{SW}_{u}$ represent the time to isolate fault $u$ and to repair fault $u$ respectively. After obtaining the nodal reliability indices, EENT can be calculated by (5).
\begin{equation}
E E N T=\frac{u_{d}}{8760} \sum_{k \in \Psi_{N}^{WT}} T I D_{k} R_{k}
\end{equation}\par
Therefore, the calculation of $m_i^u$ and $n_i^u$ is the key point of RA. It requires a clear correspondence between outage events and these variables, which will be discussed in depth in Sections III and IV.
\vspace{-1em}
\subsection{Assumptions}
The following assumptions are adopted for tractability. 
\begin{itemize}
\item[1)]
The ECS is modeled as a graph/network consisting of nodes denoting offshore substations or WTs and the cable connections between nodes. The coordinates of the substations and WTs have been determined, and the ECS operates radially to avoid higher fault currents in loop operation \cite{lee2009a}.
\item[2)]
The power generation information of WTs is known, and the DC power flow model is adopted for ECS in both normal and fault states \cite{shen2023a}.
\item[3)]
The contingency set consists of cable faults and WT faults. Given the low failure rate of submarine cables, it is unlikely that multiple cables will fail simultaneously. Hence, we assume cable contingency set contains only single cable outages.
\item[4)]
The WTs in the ECS are equipped with essential protection devices and switches that can automatically disconnect a faulty WT, thereby preventing its impact on other WTs, thus the network reconfiguration is unnecessary in a WT fault.
\end{itemize}

\section{Smart Switch Configuration and Its Reliability Assessment}
This section presents the proposed smart switch configuration, in which the post-fault network recovery in ECS is possible, and describes how the system can reconfigure the network after a cable fault occurs, with a simple example. We then establish an accurate RA model based on MILP for ECS with the smart configuration. The model not only calculates reliability indices, but also generates optimal reconfiguration plans under fault scenarios to restore the power collection of WTs as much as possible. Therefore, it minimizes the economic loss of the OWFs in fault scenarios as well (if the network reconfiguration strategies are implemented). Furthermore, this section introduces potential extended applications of the model to the ECS planning problem.\par
\subsection{Post-Fault Network Recovery Switch Configuration} 
We propose a smart ECS switch configuration that enables network reconfiguration after faults. This configuration involves equipping each feeder with a circuit breaker (CB) near the substation side, which can respond to persistent cable faults at any location in the system. Additionally, isolation switches (SWs) are installed at both ends of all cables to realize local fault isolation. \par

\begin{figure}[!htbp]
\vspace{-1em}
\centering
\includegraphics[width=3.39in]{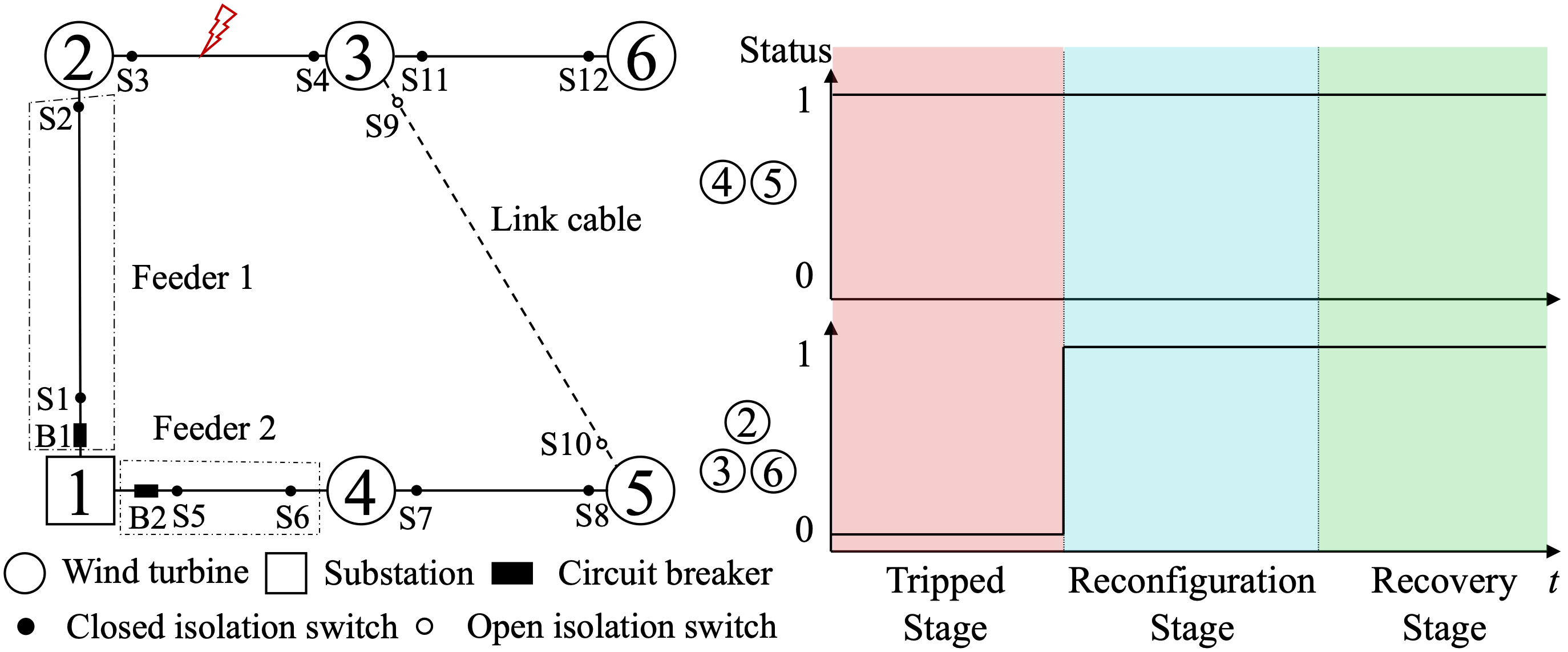}
\vspace{-0.3em}
\caption{Illustrative example of a simple system. }
\label{Simple ECS}
\vspace{-0.5em}
\end{figure}

The schematic diagram of a simple ECS with the post-fault network recovery switch configuration is shown in Fig. \ref{Simple ECS}. This system comprises one offshore substation node and five WT nodes. Solid lines represent connected cables under normal operation.
WTs 2, 3 and 6 transmit power to the substation via feeder 1, while WTs 4 and 5 transmit power via feeder 2. The dotted line between WT 3 and 5 (denoted as cable 3-5 henceforth) is the link cable, which is disconnected under normal operation. Thus, this ECS has a ring/loop structure but operates radially, corresponding to Assumption 1). \par

This system is taken as an example to demonstrate the network reconfiguration process after the cable failure. Let us assume there is a persistent fault on cable 2-3.
Initially, Breaker B1 on feeder 1 trips automatically, leading to the failure of WTs 2, 3, and 6 to transmit power to the substation. After the duration $\tau^{SW}$, SWs S3 and S4 on the faulty cable are disconnected to isolate the fault locally. Once the fault is isolated, B1 is reclosed, and WT 2 resumes power transmission. Next, SWs S9 and S10 on the link cable are closed, and WTs 3 and 6 resupply power to the offshore substation through feeder 2. The network reconfiguration is completed. After that, it takes the duration $\tau^{RP}$ to eliminate the fault on cable 2-3. Then, S3 and S4 are closed, S9 and S10 are opened, and the network restores the original normal operation state. The timeline is: CB tripping, fault isolation and WTs resupply, and normal operation restoration after fault clearance. Therefore, the entire process can be divided into three stages: \textit{tripped stage}, \textit{(network) reconfiguration stage}, and \textit{recovery stage}. Fig. \ref{Simple ECS} displays the operating status of WTs in all three stages.\par
Table \ref{table_1} summarizes the impact of all potential single cable outages in the illustrative example. It can be inferred from the table that, during a single cable fault, the WTs in the same feeder as the faulty cable are affected in the \textit{tripped stage}, while other WTs are not. Most affected WTs could transmit power via another feeder after reconfiguration in the \textit{reconfiguration stage}, but not always. The fifth row reflects if cable 3-6 fails, WT 6 cannot connect to the substation through other cables, and its power transmission can not be restored until the \textit{recovery stage}\textit.\par
\begin{table*}[htbp]
\begin{center}
\caption{Impact of All Potential Single Cable Outages in Illustrative Example }
\label{table_1}
\begin{tabular}{@{}cccccccc@{}}
\toprule
Fault events & \multicolumn{2}{c}{Actions after failures}           & \multicolumn{5}{c}{Duration of power supply interruption}                        \\ \midrule 
Faulted cable &
  \begin{tabular}[c]{@{}c@{}}Switch operation \\ for fault isolation\end{tabular} &
  \begin{tabular}[c]{@{}c@{}}Switch operation\\ for reconfiguration\end{tabular} &
 Node 2 &
 Node 3 &
 Node 4 &
 Node 5 &
 Node 6 \\ \midrule
1-2          & Open B1, open S1, S2   & Close B1, close S9, S10 & $\tau^{SW}$ & $\tau^{SW}$ & --         & --         & $\tau^{SW}$ \\
2-3          & Open B1, open S3, S4   & Close B1, close S9, S10 & $\tau^{SW}$ & $\tau^{SW}$ & --         & --         & $\tau^{SW}$ \\
3-6          & Open B1, open S11, S12 & Close B1                & $\tau^{SW}$ & $\tau^{SW}$ & --         & --         & $\tau^{SW}+\tau^{RP}$  \\
1-4          & Open B2, open S5, S6   & Close B2, close S9, S10 & --         & --         & $\tau^{SW}$ & $\tau^{SW}$ & --         \\
4-5          & Open B2, open S7, S8   & Close B2, close S9, S10 & --         & --         & $\tau^{SW}$ & $\tau^{SW}$ & --         \\ \bottomrule
\end{tabular}
\vspace{-2em}
\end{center}
\end{table*}

\subsection{The First Reliability Assessment Model} 
Considering the aforementioned network reconfiguration and fault clearance process, the first RA model for the ECS with post-fault network recovery switch configuration, denoted by \textit{RA1}, is formulated as follows:\par
\begin{equation} 
\min \quad EENT
\end{equation}
$s. t.$
\begin{equation} 
P_{ij}^{rs}=\sum_{k \in \Psi_i} P_{ki}^{rs}+P_i^{rs}, \forall i \in \Psi_N^{WT}
\end{equation} 
\begin{equation} 
\left|B_{i j}\left(\theta_j^{rs}-\theta_i^{rs}\right)-P_{i j}^{rs}\right| \leq\left(1-s_{ij}^{rs}\right) M ,\forall ij \in \Psi_C
\end{equation}
\begin{equation} 
\theta_j^{rs}=0,\quad \forall j \in \Psi_N\setminus\Psi_N^{WT}
\end{equation}
\begin{equation} 
P_{f}^{rs}=P_{br^f}^{rs}, \quad \forall f \in \Psi^{F},  br^{f} \in \Psi_C
\end{equation}
\begin{equation} 
-M s_{i j}^{rs} \leq P_{i j}^{rs} \leq M s_{i j}^{rs}, \forall i j \in \Psi_C
\end{equation}
\begin{equation} 
-P_{i j}^{C} \leq P_{i j}^{rs} \leq P_{i j}^{C}, \forall i j \in \Psi_C
\end{equation}
\begin{equation} 
P_{f}^{rs} \leq P_{f}^{C}, \forall f \in \Psi_{F}
\end{equation}
$$
\forall rs \in \Psi_C \cup\{N O\} \quad {for} \quad (7)-(13)
$$
\begin{equation} 
s_{rs}^{rs}=0
\end{equation}
\begin{equation} 
h_{k}^{f}+h_{rs}^{f}-1 \leq m_{k}^{rs}, \forall f \in \Psi_F, \forall k \in \Psi_N^{WT}
\end{equation}
\begin{equation} 
m^{rs}_k \geq n^{rs}_k, \forall k \in \Psi_{N}^{WT}
\end{equation}
\begin{equation} 
P_{k}^{rs}=P_{k}\left(1-n_{k}^{rs}\right), \forall k \in \Psi_{N}^{WT}
\end{equation}
\begin{equation} 
\sum_{i j \in \Psi_C}s_{ij}^{rs}=\sum_{k \in \Psi_{N}^{W T}}\left(1-n_{k}^{rs}\right)
\end{equation}
$$
\forall rs \in \Psi_C \quad {for} \quad (14)-(18)
$$
\begin{equation}
T I F_{k}=\sum_{rs\in \Psi_C} \lambda_{rs} m_{k}^{rs}+\lambda_k
\end{equation}
\begin{equation}
\begin{array}{c}
T I D_{k}=\sum\limits_{rs\in \Psi_C} \lambda_{rs}\left(\tau^{SW} m_{k}^{rs}+\tau^{RP} n_{k}^{rs}\right)+\lambda_k\tau^{WT}
\end{array}
\end{equation}
$$
\forall k \in \Psi_N^{WT} \quad {for} \quad (19)-(20)
$$
\begin{equation}
\begin{array}{l}
EENT=\\
\frac{u_{d}}{8760} \sum\limits_{k \in \Psi_{N}^{WT}}\sum\limits_{rs\in \Psi_C} \left(\lambda_{rs}\left(\tau^{SW} m_{k}^{rs}+\tau^{RP} n_{k}^{rs}\right)+\lambda_k\tau^{WT}\right)R_{k} 
\end{array}
\end{equation}
\begin{equation}
C_{r e l}=\alpha\cdot EENT\frac{(1+r)^t-1}{r(1+r)^t}
\end{equation}

It is worth noting that superscript $rs$ represents different scenarios. Namely, $rs \in \Psi_C$ describes a scenario where cable $rs$ (cable between node $r$ and $s$) fails, while $rs \in \{NO\}$ represents the normal operation scenario (no fault happens). \par

Among them, formulas (6)-(22) are used to describe the network reconfiguration for each cable fault scenario, thus producing $\left\{m_{k}^{rs}\right\}_{rs \in \Psi_C, k \in \Psi_{N}^{WT}}$ and $\left\{n_{k}^{rs}\right\}_{rs \in \Psi_C, k \in \Psi_{N}^{WT}}$, which can be used to calculate reliability indices as in Eq. (19)-(21).
The objective function (6), with details shown in Eq. (21), aims to minimize wind power curtailment in cable and WT failure scenarios, so that the most effective fault-handling measures can be obtained accordingly. \par

%
The DC flow model is adopted in this paper, as shown in the constraints (7) to (10). Eq. (7) is the power balance constraint. With the big-M method, constraint (8) represents the phase relationship between nodes $i$ and $j$ when cable $ij$ is connected in the fault scenario of cable $rs$. The offshore substation is set as the reference node with its voltage phase set to 0, as shown in Eq. (9). 

Constraint (11) couples the power flow on the cable with its connection state, ensuring that the power flow is zero when the cable is disconnected.  Constraints (12) and (13) limit the power flow on the cable and the feeder, respectively, to their rated capacities. Constraint (14) ensures that the faulty cable is disconnected until it is repaired.\par
When cable $rs$ fails, the CB on the feeder to which cable $rs$ belongs will trip, making the WTs connected to that feeder lose their power transmission capacity. The fault impact constraint (15) ensures that the affected WTs cannot supply power.  
Constraint (16) means the WTs that have not lost power transmission capacity due to CB tripping should maintain power supply after the network is reconfigured. That is, for any WT $k$, if $m_k^{rs}=0$, $n_k^{rs}=0$ should hold. 
Constraint (17) is the coupling constraint between the sent power of WTs $P_k^{rs}$ and the fault continuation variables $n_k^{rs}$ ($n_k^{rs}=1$ when WT $k$ cannot supply power after reconfiguration), and constraint (18) ensures that the system operates radially based on its spanning tree topology.\par

The node reliability indices TIF and TID can be obtained by Eq. (19)-(20), while the ECS's reliability index EENT is decided by Eq. (21). To quantify the reliability-related cost, we calculate the economic loss caused by the curtailed wind power due to cable and WT failures by Eq. (22). Therefore, with the proposed model, we can determine the network reconfiguration strategies of the ECS minimizing wind power curtailment due to contingencies.
\vspace{-1em}
\subsection{Potential Applicability of RA1}
In this part, we will discuss how to embed \textit{RA1} into the planning model as an explicit expression of the reliability of the ECS, whose topology could vary in the optimization process. This will assist ECS designers in achieving a balance between system economy and reliability.


One crucial step in performing RA is to solve for the cable fault impact variable $m_k^{rs}$ ($m_k^{rs}=1$ when WT $k$ is affected by the failure of cable $rs$), and Eq. (15) is proposed to address this issue.\par

If the topology of the ECS is known, as the illustrative example in Fig. 2 shows, both $h_{rs}^f$ and $h_k^f$ are parameters, which can be acquired from the system structure diagram. 

\begin{equation}
\left\{\begin{array} { l } 
{ h _ { 1 2 } ^ { 1 } = 1 , h _ { 1 2 } ^ { 2 } = 0 } \\
{ h _ { 2 3 } ^ { 1 } = 1 , h _ { 2 3 } ^ { 2 } = 0 } \\
{ h _ { 3 6 } ^ { 1 } = 1 , h _ { 3 6 } ^ { 2 } = 0 } \\
{ h _ { 1 4 } ^ { 1 } = 0 , h _ { 1 4 } ^ { 2 } = 1 } \\
{ h _ { 4 5 } ^ { 1 } = 0 , h _ { 4 5 } ^ { 2 } = 1 }
\end{array} \quad \left\{\begin{array}{l}
h_2^1=1, h_2^2=0 \\
h_3^1=1, h_3^2=0 \\
h_4^1=0, h_4^2=1 \\
h_5^1=0, h_5^2=1 \\
h_6^1=1, h_6^2=0
\end{array}\right.\right.
\end{equation}

Thus, combined with (15), $m_k^{rs}$ can be calculated as follows.
\begin{equation}
\left\{\begin{array}{l}
m_{2}^{12}=1, \,m_{2}^{23}=1, \,m_{2}^{36}=1, \,m_{2}^{14}=0, \,m_{2}^{45}=0 \\
m_{3}^{12}=1, \,m_{3}^{23}=1, \,m_{3}^{36}=1, \,m_{3}^{14}=0, \,m_{3}^{45}=0 \\
m_{4}^{12}=0, \,m_{4}^{23}=0, \,m_{4}^{36}=0, \,m_{4}^{14}=1, \,m_{4}^{45}=1 \\
m_{5}^{12}=0, \,m_{5}^{23}=0, \,m_{5}^{36}=0, \,m_{5}^{14}=1, \,m_{5}^{45}=1 \\
m_{6}^{12}=1, \,m_{6}^{23}=1, \,m_{6}^{36}=1, \,m_{6}^{14}=0, \,m_{6}^{45}=0
\end{array}\right.
\end{equation}
which are consistent with the results in Table I.\par
If the model is incorporated into the ECS planning process for RA of networks with undetermined topology, $h_{rs}^f$ and $h_k^f$ become decision variables. It's necessary to describe them with the decision variables of ECS planning. Obviously, when cable $ij$ is connected ($s_{ij}^{NO}=1$), nodes $i$ and $j$, as well as cable $ij$, are affiliated to the same feeder. This can be expressed as nonlinear constraint (25).\par
\begin{equation}
h_{ij}^f=s_{ij}^{NO}h^f_i=s_{ij}^{NO}h^f_j
\end{equation}
One can linearize constraints that contain bi-linear terms to make them more manageable by applying the big-M method (the same technique is also applied to logical constraints in later sections, but will not be emphasized). By linearizing Eq. (25) we get constraints (26)-(27). The source of the affiliation relationship is given by Eq. (28). And if cable $ij$ is disconnected under normal operation ($s_{ij}^{NO}=0$), it does not belong to any feeder, as indicated in constraint (29).\par
\begin{equation}
\begin{aligned}
\left| h_{i j}^{f}-h_{i}^{f}\right| \leq M\left(1-s_{i j}^{NO}\right), & \forall ij \in \Psi_C, \forall f \in \Psi_F
\end{aligned}
\end{equation}
\begin{equation}
\begin{aligned}
\left| h_{i j}^{f}-h_{j}^{f}\right| \leq M\left(1-s_{i j}^{NO}\right), & \forall ij \in \Psi_C, \forall f \in \Psi_F
\end{aligned}
\end{equation}
\begin{equation}
h_{br^f}^{f}=s_{br^f}^{NO},\forall f \in \Psi_F, br^f \in \Psi_C
\end{equation}
\begin{equation}
h_{ij}^{f} \leq s_{ij}^{NO},\forall f \in \Psi_F, \forall ij \in \Psi_C
\end{equation}
\begin{equation}
0 \leq h_{k}^{f} \leq 1, \quad \forall k \in \Psi_N^{WT}, \quad \forall f \in \Psi^{F} 
\end{equation}
\begin{equation}
0 \leq h_{ij}^{f} \leq 1, \quad \forall i j \in \Psi_C, \quad \forall f \in \Psi^{F}
\end{equation}
\begin{equation}
\sum_{f} h_{k}^{f} \leq 1, \quad \forall k \in \Psi_N^{WT}
\end{equation}
\begin{equation}
\sum_{f} h_{ij}^{f} \leq 1, \quad \forall i j \in \Psi_C
\end{equation}

Thus, the proposed \textit{RA1}'s potential application in the field of ECS network planning and operation can be achieved with the constraints above.
As the model is an MILP problem, its computational efficiency is determined by the number of binary variables somehow. 
To reduce this number, we set $h_{ij}^f$ and $h_k^f$ as continuous variables during modeling and add their coupling constraints with binary variables by (26)-(29). We also specify their value ranges using (30)-(33), thus $h_{ij}^f$ and $h_k^f$ can only take the value of 0 or 1. This is equivalent to making them binary variables, but the total number of binary variables in the model is reduced by $n_{f}n_{n}+n_{f}n_{c}$, where $n_{f}$, $n_{n}$, and $n_{c}$ denote the number of feeders, nodes, and cables respectively, thereby improving the computational efficiency of the model.\par
\section{ECS Reliability Assessment \\ Considering Detailed Switch Deployment}
To investigate the impact of switch configuration, i.e., the deployment of CBs and SWs, on the reliability of the ECS, and to assess whether the proposed smart switch configuration is a worthwhile investment, we extend the \textit{RA1} to consider the flexible placement of switch devices, using the virtual fault flow (VFF) method.
\vspace{-1em}
\subsection{Description of VFF}
\begin{figure*}[htbp]
    \centering
    \includegraphics[width=\textwidth]{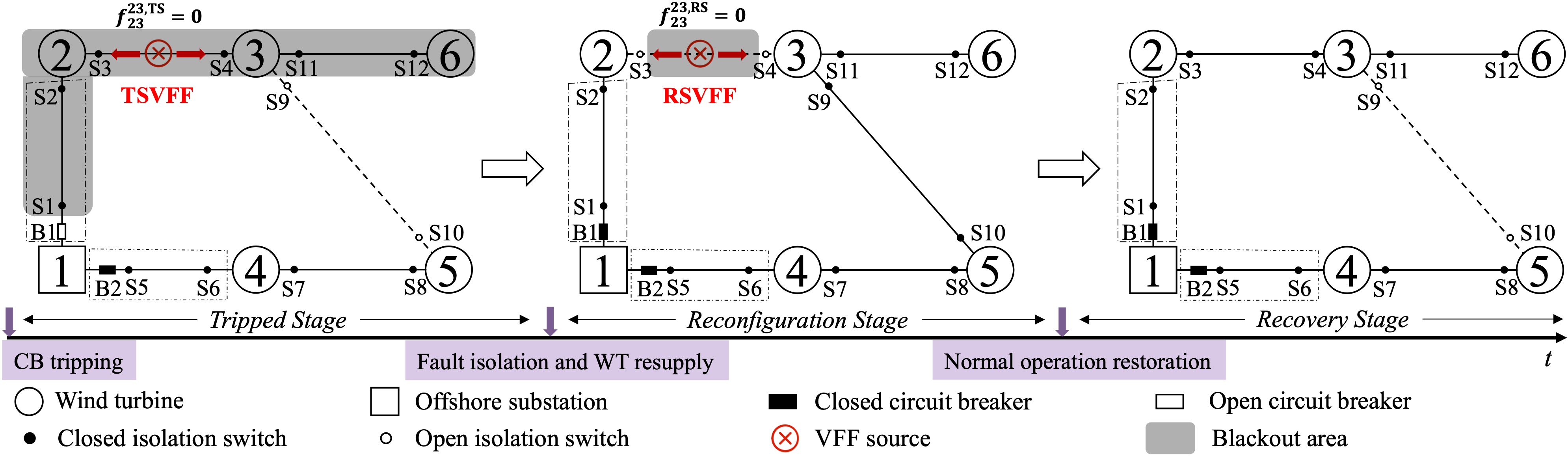}
    \caption{Description of VFF.}
    \label{VFF}
    \vspace{-1em}
\end{figure*}

VFF refers to the simulated propagation of "faulty flow" under cable fault scenarios, which enables us to evaluate fault impact ranges. Essentially, VFF simulates the isolation of the fault area by switch devices in the ECS, causing the part that the VFF flows through to experience a blackout. VFF arises from the faulted cable and can be classified into two types based on the stage of the fault: 1) \textit{tripped stage} virtual fault flow (TSVFF) and 2) \textit{reconfiguration stage} virtual fault flow (RSVFF). TSVFF can be interrupted by open SWs during normal operation and tripped CBs, while RSVFF can be interrupted by open SWs during the reconfiguration stage, as illustrated in Fig. \ref{VFF}.\par

The example in Fig. \ref{Simple ECS} will still be used for illustration. When cable 2-3 experiences a persistent fault and breaker B1 automatically trips, the system enters the \textit{tripped stage}. TSVFF originates from cable 2-3 and spreads to both sides. TSVFF is blocked by the tripped B1 when it spreads upstream. As S11 and S12 are closed, while S9 and S10 are open during normal operation, TSVFF can propagate downstream along cable 3-6 but not along cable 3-5. 
Once the switches are operated by the OWF's operator, the ECS enters the \textit{reconfiguration stage}. RSVFF arises from cable 2-3 and is restricted between S3 and S4 since the two SWs are open in this stage. The shaded areas in Fig. \ref{VFF} indicate the blackout areas caused by the TSVFF and RSVFF.

\subsection{The Second Reliability Assessment Model}

The second RA model (dentoed by \textit{RA2}), which considers flexible switch deployment, also includes TIF, TID, and EENT as reliability indices. The objective function is consistent with Eq. (6) and (21), aimed at obtaining reconfiguration strategies that result in the least wind power curtailment under various fault scenarios.  Namely, still
\begin{equation}
\min \quad EENT
\end{equation}
\par The constraints can be divided into four parts. The first two parts simulate the VFF propagation in different stages. To be more specific, the first part of constraints is about the spread of TSVFF in the \textit{tripped stage} (TS):
\begin{equation}
f_{rs}^{rs, \mathrm{TS}}=0
\end{equation}
\begin{equation}
\begin{array}{c}
\left|f_{i j}^{rs, \mathrm{TS}}-f_i^{rs, \mathrm{TS}}\right| \leq\left(1-b_{i j}^{i, rs}\right) M, \forall i j \in \Psi_I^B 
\end{array}
\end{equation}
\begin{equation}
\begin{array}{c}
\left|f_{i j}^{rs, \mathrm{TS}}-f_i^{rs, \mathrm{TS}}\right| \leq\left(1-s_{i j}^{i, \mathrm{NO}}\right) M, \forall i j \in \Psi_I^S, i j \notin \Psi_I^B 
\end{array}
\end{equation}
\begin{equation}
\begin{array}{c}
f_{i j}^{rs, \mathrm{TS}}=f_i^{rs, \mathrm{TS}}, \forall i j \notin \Psi_I^S, i j \notin \Psi_I^B 
\end{array}
\end{equation}
\begin{equation}
\begin{array}{c}
\left| f_{i j}^{rs, \mathrm{TS}}-f_j^{rs, \mathrm{TS}}\right| \leq\left(1-b_{i j}^{j, rs}\right) M, \forall i j \in \Psi_J^B 
\end{array}
\end{equation}
\begin{equation}
\begin{array}{c}
\left| f_{i j}^{rs, \mathrm{TS}}-f_j^{rs, \mathrm{TS}}\right| \leq\left(1-s_{i j}^{j, \mathrm{NO}}\right) M, \forall i j \in \Psi_J^S, i j \notin \Psi_J^B
\end{array}
\end{equation}
\begin{equation}
\begin{array}{c}
f_{i j}^{rs, \mathrm{TS}}=f_j^{rs, \mathrm{TS}}, \forall i j \notin \Psi_J^S, i j \notin \Psi_J^B
\end{array}
\end{equation}
\begin{equation}
\begin{array}{c}
0 \leq f_i^{rs, \mathrm{TS}} \leq 1, \forall i \in \Psi_N^{WT}
\end{array}
\end{equation}
\begin{equation}
\begin{array}{c}
0 \leq f_{i j}^{rs, \mathrm{TS}} \leq 1, \forall i j \in \Psi_C
\end{array}
\end{equation}
\begin{equation}
\begin{array}{c}
f_i^{rs, \mathrm{TS}}=1, \forall i \in \Psi_{N}\setminus\Psi_N^{WT}
\end{array}
\end{equation}
\begin{equation}
\begin{array}{c}
\sum\limits_{ij\in\Psi_I^B}\left|b_{i j}^{i,\mathrm{NO}}-b_{i j}^{i, rs}\right|+\sum\limits_{ij\in\Psi_J^B}\left|b_{i j}^{j,\mathrm{NO}}-b_{i j}^{j, rs}\right|\leq1
\end{array}
\end{equation}
\begin{equation}
\begin{array}{c}
m_k^{rs}=1-f_k^{rs, \mathrm{TS}}, \forall k \in \Psi^{WT}_N
\end{array}
\end{equation}
\par As mentioned earlier, TSVFF arises from the faulty cable, which is described in Eq. (35). Its propagation in ECS is impacted by CBs and SWs. Only the CBs that trip during the TS or the SWs that are disconnected in normal operation condition can interrupt the spread of TSVFF, and constraints (36)-(41) describe this logic using big-M method. To reduce the number of binary variables, the same approach as in Section III C. is adopted. The VFF variables are defined as continuous variables, and their values are limited by constraints (42)-(44). Eq. (45) emphasizes that there should be at most one CB trip action after cable fault occurs. Eq. (46) denotes the relationship between the fault impact variables and the TSVFF variables.\par
The second part of constraints is about the spread of RSVFF in the \textit{reconfiguration stage} (RS):
\begin{equation}
f_{rs}^{rs, \mathrm{RS}}=0
\end{equation}
\begin{equation}
\begin{array}{c}
\left| f_{i j}^{rs, \mathrm{RS}}-f_i^{rs, \mathrm{RS}}\right| \leq\left(1-s_{i j}^{i, rs}\right) M, \forall i j \in \Psi_I^S \\
\end{array}
\end{equation}
\begin{equation}
f_{i j}^{rs, \mathrm{RS}}=f_i^{rs, \mathrm{RS}}, \forall i j \notin \Psi_I^S 
\end{equation}
\begin{equation}
\begin{array}{c}
\left|f_{i j}^{rs, \mathrm{RS}}-f_j^{rs, \mathrm{RS}}\right|  \leq\left(1-s_{i j}^{j, rs}\right) M, \forall i j \in \Psi_J^S \\
\end{array}
\end{equation}
\begin{equation}
f_{i j}^{rs, \mathrm{RS}}=f_j^{rs, \mathrm{RS}}, \forall i j \notin \Psi_J^S 
\end{equation}
\begin{equation}
0 \leq f_i^{rs, \mathrm{RS}} \leq 1, \forall i \in \Psi^{WT}
\end{equation}
\begin{equation}
\begin{array}{c}
0 \leq f_{i j}^{rs, \mathrm{RS}} \leq 1, \forall i j \in \Psi_C \\
\end{array}
\end{equation}
\begin{equation}
\begin{array}{c}
f_i^{rs, \mathrm{RS}}=1, \forall i \in \Psi_{N}\setminus\Psi_N^{WT} \\
\end{array}
\end{equation}
\begin{equation}
n_k^{rs}=1-f_k^{rs, \mathrm{RS}}, \forall k \in \Psi^{WT}_N
\end{equation}
$$
\forall rs \in \Psi_C \quad {for} \quad (35)-(55)
$$
\par Similarly, RSVFF arises from the faulty cable, as described in Eq. (47).
Its propagation within the ECS is impacted by SWs. Only the SWs that are disconnected during the RS could interrupt the spread of RSVFF. This logic is described by constraints (48)-(51). RSVFF variables are defined as continuous variables with their values restricted by constraints (52)-(54). Eq. (55) denotes the relationship between the fault continuation variables and VFF variables.\par
The operational constraints under fault scenarios are formulated in the third part:
$$
(7)-(13), \forall rs \in \Psi_C \cup\{NO\} \quad (16)-(18), \forall rs \in \Psi_C
$$
\begin{equation}
s^{rs}_{ij}=1, \forall ij \notin \Psi_I^S \ \text{and} \ ij \notin \Psi_J^S
\end{equation}
\begin{equation}
s^{rs}_{ij}=s^{i,rs}_{ij}, \forall ij \in \Psi_I^S \ \text{and} \ ij \notin \Psi_J^S
\end{equation}
\begin{equation}
s^{rs}_{ij}=s^{j,rs}_{ij}, \forall ij \in \Psi_J^S \ \text{and} \ ij \notin \Psi_I^S
\end{equation}
\begin{equation}
\left\{\begin{array}{l}
s^{rs}_{ij}\leq s^{i,rs}_{ij}\\
s^{rs}_{ij}\leq s^{j,rs}_{ij}\\
s^{rs}_{ij}\geq s^{i,rs}_{ij}+s^{j,rs}_{ij}-1
\end{array}\right. , \forall ij \in \Psi_I^S \ \text{and} \ ij \in \Psi_J^S
\end{equation}
which are similar to \textit{RA1}. But due to the consideration of detailed switch deployment, the connectivity status of the cable depends on the connectivity status of its corresponding switches, as described in (56)-(59). \par
The fourth part of constraints calculates the reliability indices and comprehensive benefits of switch configurations:
$$
(19)-(22)
$$
\begin{equation}
\begin{array}{l}
V=\\
\alpha\left(EENT_0-EENT\right)\frac{(1+r)^t-1}{r(1+r)^t}-\left(c_{CB}n_{CB}+c_{SW}n_{SW}\right)
\end{array}
\end{equation}
The overall benefit of switch configuration, denoted by $V$ in Eq. (60), is defined as the benefit of reliability improvement ($\alpha\left(EENT_0-EENT\right)\frac{(1+r)^t-1}{r(1+r)^t}$) minus the cost of purchasing and installing all switch devices ($c_{CB}n_{CB}+c_{SW}n_{SW}$). Obviously, the benefit brought by reliability improvement is equal to the difference between the blackout costs of the no-switch configuration ($EENT_0$) and the current configuration ($EENT$). The present value of the overall benefit of switch configuration is considered by multiplying a coefficient $\frac{(1+r)^t-1}{r(1+r)^t}$.\par
Note that, both \textit{RA1} and \textit{RA2} are formulated and transformed into MILP forms, which can be easily solved by modern branch-and-cut solvers.

\section{Case Study}
The proposed \textit{RA1} and \textit{RA2} are applied and verified in this section. 
Firstly, we study the influence of post-fault network reconfiguration as well as the system operating state on reliability at the Ormonde OWF. Next, we apply \textit{RA1} to examine the impact of ECS topology on reliability at the Hornsea One Centre OWF. Finally, we assess the reliability of the Beatrice OWF equipped with six different switch configurations with \textit{RA2}. The information of three real OWFs is obtained from \cite{kisorca}. For ease of reproducibility, the detailed data used for numerical experiments can be found in \cite{Ding2023}.
Simulations have been implemented on a laptop PC with an Intel Core i5 processor using Gurobi 10.0.0. The optimality tolerance is set to 0 so that all cases are solved to global optimality.\par

\subsection{Effect of Post-Fault Network Reconfiguration on Reliability}
\begin{figure}[htbp]
\centering
\includegraphics[width=3.39in]{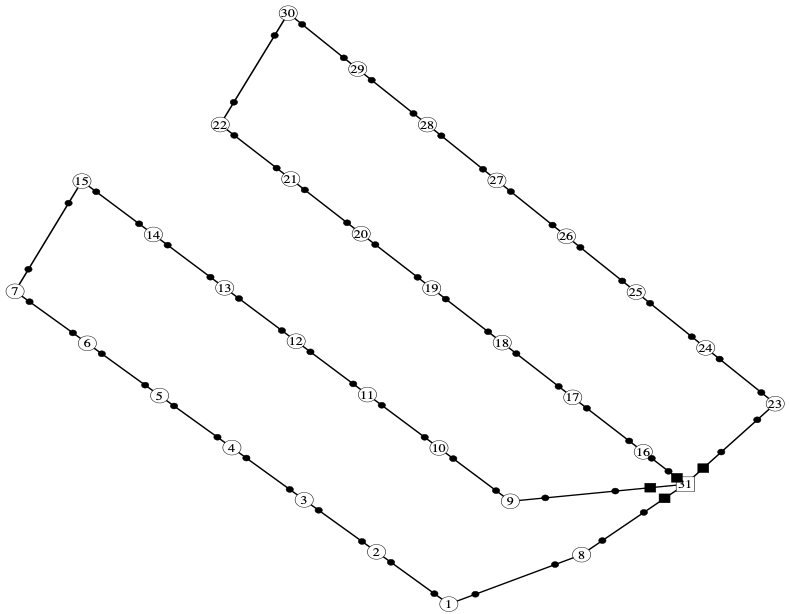}
\vspace{-0.5em}
\caption{ECS of the Ormonde OWF. }
\label{Ormonde}
\vspace{-2em}
\end{figure}

Here, we research the effect of post-fault network reconfiguration on reliability through the Ormonde OWF shown in Fig. \ref{Ormonde}. We adopt two approaches to obtain various operating states for the radial and ring ECS, respectively. One approach involves sequentially removing two cables in Fig. \ref{Ormonde} to form the radial topology and treating it as the radial-ECS normal operating state. The other approach obtains the operating states conforming to the assumption of radial operation by sequentially selecting two specific cables in Fig. \ref{Ormonde} as the normally disconnected cables (link cables).
%
The \textit{RA1} is applied to every operational state, and Table \ref{EffectofReconfiguration} shows the comparison of the system's reliability index with and without network reconfiguration.\par

\vspace{-1em}

\begin{table}[htbp]
\begin{center}
\caption{Influence of Network Reconfiguration on Reliability}
\vspace{-2em}
\label{EffectofReconfiguration}
\setlength{\tabcolsep}{3.7mm}{
\begin{tabular}{@{}cccc@{}}
\toprule
\multicolumn{2}{c}{\begin{tabular}[c]{@{}c@{}}Radial ECS\\ (without network reconfiguration)\end{tabular}} &
  \multicolumn{2}{c}{\begin{tabular}[c]{@{}c@{}}Ring ECS\\ (with network reconfiguration)\end{tabular}} \\ \midrule
\multirow{3}{*}{\begin{tabular}[c]{@{}c@{}}Removed\\ cables\\ $i$-$j$\end{tabular}} &
  \multirow{3}{*}{\begin{tabular}[c]{@{}c@{}}$EENT$\\ (MWh/yr)\end{tabular}} &
  \multirow{3}{*}{\begin{tabular}[c]{@{}c@{}}Link\\ cables\\  $i$-$j$\end{tabular}} &
  \multirow{3}{*}{\begin{tabular}[c]{@{}c@{}}$EENT$\\ (MWh/yr)\end{tabular}} \\
                    &                      &                     &                    \\
                    &                      &                     &                    \\ \midrule
9-10, 16-17                 & 14117.80               & 9-10, 16-17                 & 149.72              \\
10-11, 17-18                 & 12572.70                & 10-11, 17-18                 & 139.42               \\
11-12, 18-19                 & 11265.70               & 11-12, 18-19                 & 130.79             \\
12-13, 19-20                 & 10196.30                & 12-13, 19-20                 & 123.80             \\
13-14, 20-21                & 9364.78                & 13-14, 20-21                & 118.45             \\
14-15, 21-22                & 8771.11                & 14-15, 21-22                & 114.75             \\
\textbf{7-15, 22-30}      & \textbf{8413.37}      & \textbf{7-15, 22-30}      & \textbf{110.84}     \\
6-7, 29-30               & 8430.29                & 6-7, 29-30               & 112.51             \\
5-6, 28-29               & 8683.28                & 5-6, 28-29               & 114.21             \\
4-5, 27-28                & 9174.13                & 4-5, 27-28                & 117.56             \\
3-4, 26-27                 & 9902.67                & 3-4, 26-27                 & 122.54             \\
2-3, 25-26                 & 10869.20                & 2-3, 25-26                 & 129.18               \\
1-2, 24-25                 & 12073.60                & 1-2, 24-25                 & 137.47               \\
1-8, 23-24                 & 13516.00                & 1-8, 23-24                 & 147.29               \\\bottomrule
\end{tabular}
}
\end{center}
\end{table}
\vspace{-1.5em}
%
%
It is evident from Table \ref{EffectofReconfiguration} that the ECS exhibits varying
reliability levels depending on the operating states. The proposed
model is able to identify the most reliable operational state for both radial and ring ECS.
The reliability index comparison demonstrates that the ring ECS and the implementation of network reconfiguration lead to a more reliable OWF. By comparing the reliability results for each type of ECS, it can be observed that the highest reliability is achieved when cables 7-15 and 22-30 are chosen as the ``removed cables'' or ``link cables'', while the lowest reliability is achieved when cables 9-10 and 16-17 are selected as the ``removed cables'' or ``link cables''.\par
Based on Table \ref{EffectofReconfiguration} and Fig. \ref{Ormonde}, we can conclude that if WTs are more evenly distributed on different feeders under normal operation, we are more likely to achieve a lower reliability index EENT, namely, a more reliable ECS.\par
\vspace{-1em}
\subsection{Effect of Topology on Reliability}

\begin{figure}[htbp]
\centering
\includegraphics[width=3.39in]{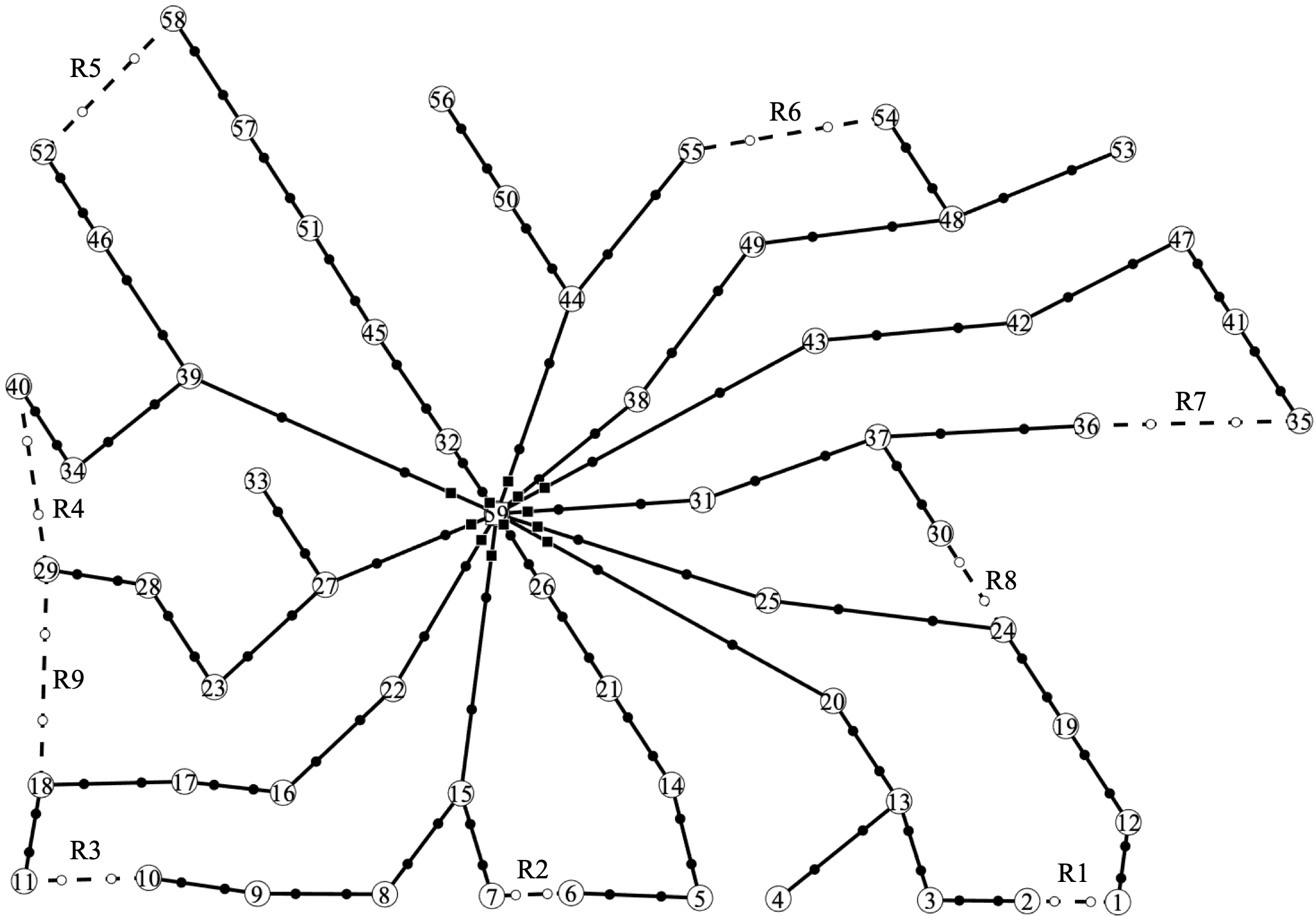}
\vspace{-0.5em}
\caption{Hornsea One Centre OWF with link cables. }
\vspace{-0.5em}
\label{Hornsea}
\end{figure}

%
In this subsection, we investigate the impact of the ECS topology on reliability indices by adding link cables to the system. To ensure experimental uniformity, we assess the reliability of different topologies under the most reliable operational states, based on the conclusions from the previous subsection. The research is conducted on the Hornsea One Centre OWF shown in Fig. \ref{Hornsea}, which contains nine link cables labeled as R1-R9.  We design ten interrelated but distinct cases, described as follows. Case I represents a radial ECS without any link cables. From Case II to X, one link cable is added to the previous case in the order of R1-R9. As such, Case X contains all nine link cables. \par

To validate the feasibility of the \textit{RA1}, we compare it with the Sequential Monte-Carlo Simulation (SMCS) method.
Tables \ref{ANALYTICAL METHOD} and \ref{SMC METHOD} summarize the RA results of ten cases based on the \textit{RA1} and SMCS method, respectively. We can see that the RA results of the two methods are basically consistent, but the proposed \textit{RA1} is more efficient, with an average computation time of 1.1s, compared to 185.7s for the SMCS. The proposed \textit{RA1} significantly outperforms the SMCS method in terms of solution speed.\par
\begin{table}[htbp]
\vspace{-1em}
\begin{center}
\caption{System Reliability Assessment with \\ Proposed Method}
\vspace{-1em}
\label{ANALYTICAL METHOD}
\setlength{\tabcolsep}{3mm}{

\begin{tabular}{@{}cccc@{}}
\toprule
Case & $EENT$(MWh) & $C_{rel}$ ($10^7$\$) & Solving time(s) \\ \midrule
I    & 38574.50  & 9.614          & 1.15            \\
II   & 31501.80  & 7.852          & 0.79            \\
III  & 27631.68  & 6.887          & 0.86            \\
IV   & 22861.40  & 5.698          & 0.87            \\
V    & 18273.00  & 4.554          & 0.95            \\
VI   & 14567.60  & 3.631          & 0.84            \\
VII  & 9677.28   & 2.412          & 0.75            \\
VIII & 2576.12   & 0.642          & 0.90            \\
IX   & 2018.42   & 0.503          & 2.01            \\
X    & 2018.42   & 0.503          & 2.33            \\ \bottomrule
\end{tabular}
}
\vspace{-1.5em}
\end{center}
\end{table}
\begin{table}[htbp]
\begin{center}
\caption{System Reliability Assessment with \\ Sequential Monte-Carlo Simulation Method}
\vspace{-1em}
\label{SMC METHOD}
\setlength{\tabcolsep}{3mm}{
\begin{tabular}{@{}cccc@{}}
\toprule
Case & $EENT$(MWh) & $C_{rel}$ ($10^7$\$) & Solving time(s) \\ \midrule
I    & 38611.56  & 9.624          & 189.54          \\
II   & 31509.90  & 7.854          & 208.95          \\
III  & 27634.40  & 6.888          & 167.74          \\
IV   & 22884.10  & 5.704          & 187.12          \\
V    & 18274.84  & 4.555          & 168.01          \\
VI   & 14567.98  & 3.631          & 184.47          \\
VII  & 9683.39   & 2.414          & 166.52          \\
VIII & 2580.85   & 0.643          & 213.21          \\
IX   & 2043.71   & 0.509          & 189.80          \\
X    & 2022.79   & 0.504          & 181.17          \\ \bottomrule
\end{tabular}
}
\end{center}
\vspace{-3em}
\end{table}

It is clear that the construction topology has a significant impact on system reliability. With the link cables added from Case I to X, the ECS transforms from a simple radial topology to a complex ring topology, and the reliability improves accordingly. Case X has an EENT of 2018.42 MWh/year, which is only 5.2\% of Case I. This is due to the fact that, although link cables do not participate in normal operation, they can be utilized to realize network reconfiguration after a sustained fault, thus allowing some of the fault-affected WTs to resupply power in the reconfiguration stage. This is reflected by the nodal reliability indices for Case I and Case X shown in Table \ref{NodalRel}.\par 
\vspace{-1em}
\begin{table}[htbp]
\begin{center}
\caption{Nodal Reliability Indices Comparison of Case I and Case X}
\vspace{-1em}
\label{NodalRel}
\setlength{\tabcolsep}{5mm}{
\begin{tabular}{@{}ccccc@{}}
\toprule
\multirow{2}{*}{Node} & \multicolumn{2}{c}{$TIF$(times/year)} & \multicolumn{2}{c}{$TID$(hours/year)} \\ \cmidrule(l){2-5}
                      & Case I           & Case X           & Case I           & Case X           \\ \midrule
38                    & 0.506            & 0.506            & 151.234          & 3.704            \\
48                    & 0.506            & 0.506            & 472.718          & 3.704            \\
49                    & 0.506            & 0.506            & 308.728          & 3.704            \\
53                    & 0.506            & 0.506            & 623.202          & 154.188          \\
54                    & 0.506            & 0.506            & 571.789          & 3.704            \\ \bottomrule
\end{tabular}
}
\end{center}
\vspace{-1.5em}
\end{table}
Table \ref{NodalRel} compares the reliability metrics of some WTs in Case I and Case X, where the difference between the two cases is the presence or absence of link cables that support reconfiguration under fault scenarios. For example, if cable 38-59 fails, in Case I, the five WTs connected to this cable could not transmit power until the fault is fully cleared. However, in Case X, the power generated by these WTs can be transmitted to the substation via the link cable R6 as its capacity allows, thus greatly increasing the ECS's ability to reduce wind power curtailment. As shown in Table \ref{NodalRel}, investing in link cables does not influence TIF, and the main benefit is to reduce TID by allowing for reconfiguration after cable faults.
\par

It is worth noting that while laying new link cables in the ECS usually significantly improves reliability, there are also cases where it does not. This diminishing marginal utility is reflected in the last three rows of Table \ref{ANALYTICAL METHOD}. Laying R8 only slightly improves reliability, reducing EENT by only 557.7 MWh. Moreover, installing R9 makes no difference to reliability. Fundamentally, this is due to the fact that the existing link cables can support the network reconfiguration effectively. Laying R9 only increases investment without changing the optimal reconfiguration strategies.
The cost of laying marine cables can be expensive in actual projects. Therefore, when planning an ECS that does not require high reliability, it is necessary to strike a balance between investment cost and reliability-related cost to optimize the total cost of the OWF. \par

\vspace{-1em}
\subsection{Effect of Switch Configuration on Reliability}
\begin{figure}[htbp]
\centering
\includegraphics[width=3.39in]{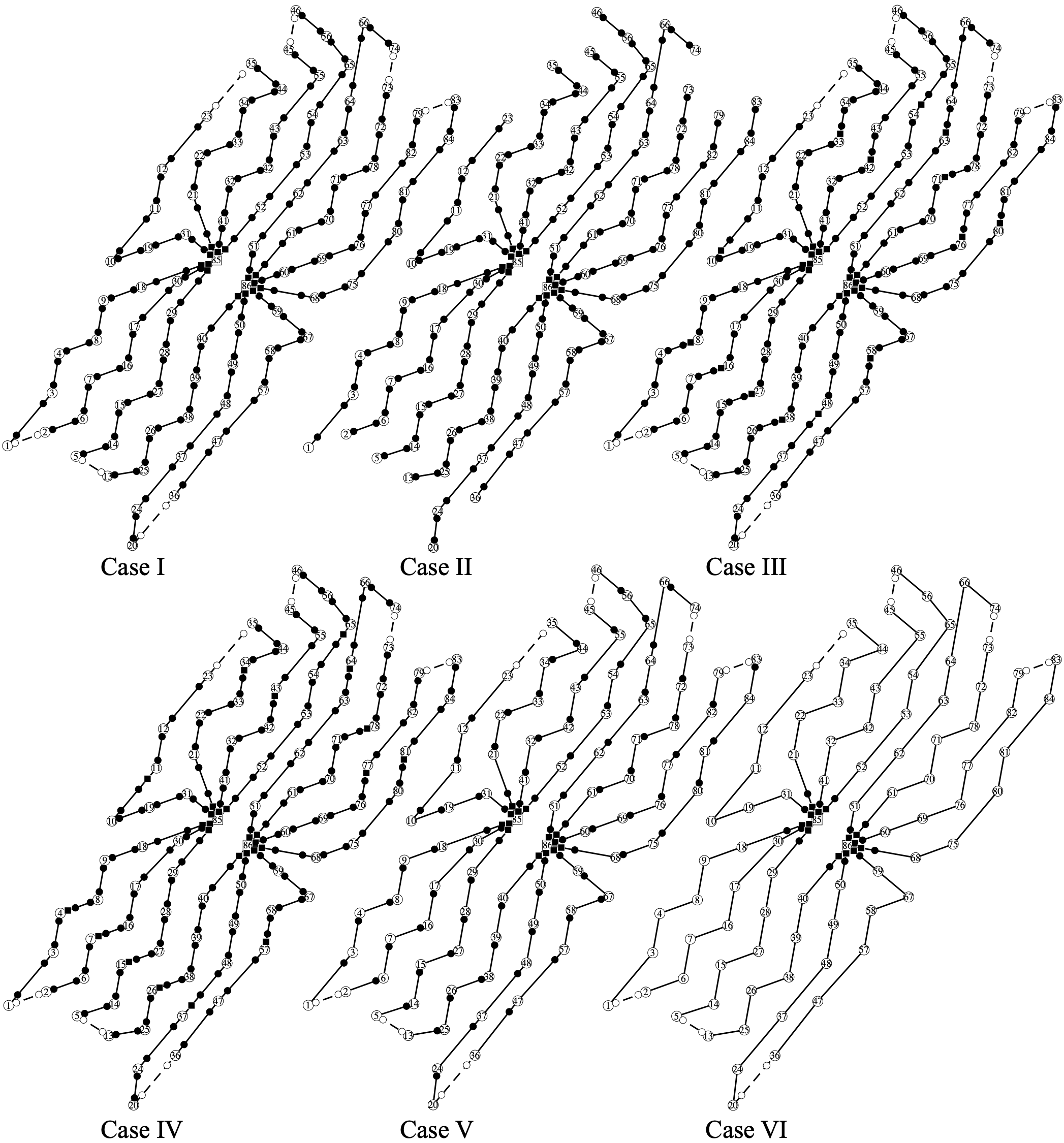}
\vspace{-0.5em}
\caption{Beatrice OWF with six switch configurations. }
\label{Beatrice}
\end{figure}

This subsection aims to verify another ECS RA model, the \textit{RA2}, considering flexible switch configurations proposed in Section IV. The proposed \textit{RA2} is highly flexible and can be applied to ECS with multiple substations. To evaluate the impact of switch configuration, we assess the reliability indices and comprehensive benefits for 6 cases on Beatrice OWF, as illustrated in Table \ref{CasesDescri} and Fig. \ref{Beatrice}.\par

\begin{table}[htbp]
\vspace{-1em}
\begin{center}
\caption{Details of Cases with Different Switch Configurations}
\label{CasesDescri}
\vspace{-1em}
\setlength{\tabcolsep}{0.5mm}
\renewcommand{\arraystretch}{1.1} 
{
\begin{tabular}{@{}cccc@{}}
\toprule
Case & Link cables & Deployment of CBs & Deployment of SWs \\ \midrule
I & \checkmark & Upstream of feeders & Both ends of all cables \\
\\
II &  & Upstream of feeders & Both ends of all cables \\
\\
III & \checkmark & \begin{tabular}[c]{@{}c@{}}Upstream of feeders and\\ upstream of selected cables\end{tabular} & Both ends of all cables \\
IV & \checkmark & \begin{tabular}[c]{@{}c@{}}Upstream of feeders and\\ downstream of selected cables\end{tabular} & Both ends of all cables \\
V & \checkmark & Upstream of feeders & \begin{tabular}[c]{@{}c@{}}Upstream of all cables and\\ both ends of link cables\end{tabular} \\
VI & \checkmark & Upstream of feeders & \begin{tabular}[c]{@{}c@{}}Upstream of feeders and\\ both ends of link cables\end{tabular} \\ \bottomrule
\end{tabular}
}
\end{center}
\vspace{-1.5em}
\end{table}

\vspace{-1em}
\vspace{-0.5em}

Results are presented in Table \ref{Switch Config}. Additionally, $EENT_0$ (as shown in Eq. (60)) of the base case without any switch device, is also calculated. When VFF propagation is not blocked by any switch device, any fault would lead to a system-wide power outage. This causes a significant curtailment, amounting to 1,330,673 MWh or 36\% of the annual power generation. Based on the results, the following conclusions can be drawn:\\
1) Comparing the base case and Cases I-VI, installing switches in ECS greatly improves the system's reliability, resulting in significant benefits. Even the installation of CBs and SWs only on feeders (as in Case VI) greatly reduces power curtailment.\\
2) With the same switch configuration, link cables can facilitate system reconfiguration, as shown by the comparison between Cases I and II, leading to further reductions in power curtailment.\\
3) Comparing Cases I and III/IV, installing sectional CBs reduces the TSVFF propagation range and the number of affected WTs in TS, improving reliability. And placing sectional CBs upstream provides greater benefits. But the installation of sectional CBs is a bit less economical due to the high cost.\\
4) Comparing Cases I and V/VI, equipping SWs on both sides of cables reduces the propagation range of RSVFF, increases the number of WTs that can recover power supply in RS, and improves system reliability.\par
\begin{table}[htbp]
\vspace{-1em}
\begin{center}
\caption{Reliability Indices and Comprehensive Benefits Comparison}
\label{Switch Config}
\vspace{-1em}
\setlength{\tabcolsep}{1.6mm}{
\begin{tabular}{@{}ccccccc@{}}
\toprule
Case                                                 & I                           & II                          & III                         & IV                          & V                           & VI                          \\ \midrule
\begin{tabular}[c]{@{}c@{}}$EENT$\\ (MWh)\end{tabular} & 433.56                      & 56401.31                    & \textbf{351.38}                      & 377.40                      & 16218.68                    & 95144.24                    \\
\begin{tabular}[c]{@{}c@{}}$V$\\ (M$\$$)\end{tabular}     & \multicolumn{1}{l}{\textbf{2262.74}} & \multicolumn{1}{l}{2167.59} & \multicolumn{1}{l}{2262.25} & \multicolumn{1}{l}{2262.21} & \multicolumn{1}{l}{2236.70} & \multicolumn{1}{l}{2103.02} \\ \bottomrule
\end{tabular}
}
\end{center}
\vspace{-1.5em}
\end{table}
We can observe that the deployment of switch devices has a profound impact on ECS's reliability.
Case III is the most reliable, while Case I has the highest comprehensive benefit. Hence, the proposed switch configuration can be a worthwhile investment for wind farm operators to consider.
\vspace{-0.5em}
\section{Discussions on Models}

In the \textit{RA1}, the number of binary variables, continuous variables, and constraints is $2n_n^2+n_cn_n+2n_n+n_c$, $2n_n^2+n_cn_n+n_fn_n+n_c+n_f+4n_n+1$, and $4n_n^2+3n_cn_n+2n_fn_n+4n_n+2$.
In the \textit{RA2}, the number of binary variables, continuous variables, and constraints is $2n_n^2+2n_{CB}n_n+2n_{SW}n_n+n_cn_n$, $4n_n^2+3n_cn_n+n_fn_n$, and $7n_n^2+10n_cn_n+2n_fn_n+9n_n+3$.\par

Our repeated tests show both models can obtain the optimal solution quickly within a few seconds. The \textit{RA1} is easier to solve and is suitable for assessment of the proposed switch configuration. The \textit{RA2} considers more details, resulting in more variables and constraints. It is recommended for analyzing the reliability of other configurations.\par

For large OWFs, more fault scenarios might need to be considered. To this end, the RA problem can be simplified and accelerated from two aspects without sacrificing accuracy. Firstly, most OWFs have radial or single/double-sided ring ECS, meaning that faults occurring on each WT string/ring will not affect other parts of the system. Hence, the ECS can be separated according to its topology, then RA for each ``independent sub-system'' can be conducted in parallel. Secondly, since the constraints of each fault scenario are naturally decoupled, i.e., without common variables, the RA problem can be further decomposed into several sub-problems according to the scenario, and thus can also be solved in parallel. Reliability metrics can be further calculated by quantifying the wind power curtailment for each sub-system under different fault scenarios.

\vspace{-0.5em}
\section{Conclusion}

RA serves as a valuable reference for the design of OWF's ECS.
In this paper, we propose a smart switch configuration in ECS, which offers network reconfiguration capability at a lower cost. Correspondingly, a RA method is presented with verified better performance against the sequential Monte-Carlo simulation method. Additionally, to evaluate different switch deployment strategies, we establish another RA model considering the detailed placement of CBs and SWs in ECS, and we show through testing that the proposed configuration offers the highest overall benefit.\par
Numerical results based on several real OWFs indicate investing in link cables usually improves reliability. But the law of diminishing marginal utility exerts itself as the number of link cables increases to a certain extent. Once the ECS topology is determined, the system's reliability depends on the switch configuration and operation state, i.e. how switch devices are placed and how evenly WTs are distributed. The application of post-fault reconfiguration strategies contributes largely to enhancing reliability. Furthermore, it is worth highlighting that by linearization, both RA models are transformed into MILP, which can be easily solved by branch-and-cut solvers.\par
For future research, it would be worthwhile to investigate the optimal planning problem to optimize the switch configuration that satisfies reliability requirements of the ECS. Furthermore, considering the proposed RA models have the potential applicability to be embeded in other optimization models of OWF as well, their incorporation (with other models), for the planning and operation of OWF, is another promising area for in-depth research.\par
\vspace{-1em}


%





\ifCLASSOPTIONcaptionsoff
  \newpage
\fi



%


\bibliographystyle{IEEEtran}
\bibliography{references.bib}

%








\end{document}